\documentclass{osa-article}

\journal{oe}

\articletype{Research Article}

\begin{document}

\title{Quantum optical frequency up-conversion for polarisation entangled qubits:\\towards interconnected quantum information devices}

\author{Florian Kaiser,\authormark{1,2,*} Panagiotis Vergyris,\authormark{1,3} Anthony Martin,\authormark{1,4} Djeylan Aktas,\authormark{1,5} Marc P. De Micheli,\authormark{1,6} Olivier Alibart,\authormark{1} and S\'ebastien Tanzilli\authormark{1}}

\address{\authormark{1}Universit\'e C\^ote d'Azur, CNRS, Institut de Physique de Nice (INPHYNI), UMR 7010, Parc Valrose, 06108 Nice Cedex 2, France\\
\authormark{2}Currently with the 3rd Institute of Physics, University of Stuttgart and Institute for Quantum Science and Technology IQST, 70569 Stuttgart, Germany\\
\authormark{3}Currently with Center for Life Nano Science$@$Sapienza, Istituto Italiano di Tecnologia, 00161 Rome, Italy\\
\authormark{4}Currently with the Group of Applied Physics, University of Geneva, Switzerland\\
\authormark{5}Currently with the Quantum Engineering Technology Labs, H. H. Wills Physics Laboratory and Department of Electrical and Electronic Engineering, University of Bristol, Bristol BS8 1FD, UK\\
\authormark{6}Deceased on July 10, 2019.}

\email{\authormark{*}f.kaiser@pi3.uni-stuttgart.de} 



\begin{abstract}
Realising a global quantum network requires combining individual strengths of different quantum systems to perform universal tasks, notably using flying and stationary qubits.
However, transferring coherently quantum information between different systems is challenging as they usually feature different properties, notably in terms of operation wavelength and wavepacket.
To circumvent this problem for quantum photonics systems, we demonstrate a polarisation-preserving quantum frequency conversion device in which telecom wavelength photons are converted to the near infrared, at which a variety of quantum memories operate. Our device is essentially free of noise which we demonstrate through near perfect single photon state transfer tomography and observation of high-fidelity entanglement after conversion. In addition, our guided-wave setup is robust, compact, and easily adaptable to other wavelengths.
This approach therefore represents a major building block towards advantageously connecting quantum information systems based on light and matter.
\end{abstract}

\section{Introduction}
Quantum technologies have the potential to revolutionise the way information is processed and communicated~\cite{Mohseni_commercialize_2017,Riedel_flagship_2017}.
Quantum computers and simulators should permit to solve hard computational tasks and simulate complex systems, respectively more efficiently than classical ones~\cite{Steane_QC_1998,Ladd_QC_2010}.
In quantum metrology, novel sensors achieve performances that are far beyond the capabilities of their classical counterparts~\cite{Giovannetti_qmet_2011,Degen_sensing_2017}. Quantum communication should bring absolute security in data exchange~\cite{Gisin_QKD_2002}.
The future very likely lies in making those quantum technological pillars compatible with each other to combine individual system advantages.
Although multi-functional quantum systems have already been demonstrated~\cite{Barz_blind_2012,Fitzsimons_blind_2017}, the usual situation today is that specific technologies are tailored for each application.
One major obstacle in connecting various quantum systems lies in the wavelength discrepancy between different systems.
For example, photons are generally preferred for quantum communication purposes~\cite{Valivarthi_tele_2016,Sun_tele_2016,Yin_DIQKD_2016}, and minimal loss in fibre networks is obtained at telecom wavelengths $(\lambda \sim 1.55\rm\,\mu m)$. On the other hand, quantum computation, storage, and metrology tasks are usually performed with matter based systems that interact with wavelengths ranging from the visible to the near infrared band $(\lambda \sim 600 - 900\rm\,nm)$~\cite{Lvovsky_memory_2009,Felix_memory_2013,Khabat_memory_2016}.

To bridge this gap, the solution lies in \textit{quantum interfaces} able to coherently convert photons back and forth between the different wavelength bands~\cite{Tanzilli_Interface}.
Such wavelength conversion interfaces are compatible with essentially all photonic observables, \textit{e.g.} energy-time~\cite{Tanzilli_Interface,Lenhard_2017_q_down_conversion_interface}, time-bin~\cite{VanDevender_upenergytime_2007,Dudin_Conversion_2010,Ikuta_2011,Maring_QFC_2017}, orbital angular momentum~\cite{Zhou_2016_OAM_interface1,Zhou_2016_OAM_interface2}, squeezed states of light~\cite{Baune_2016_Q_interface}, and polarisation~\cite{ramelow_polarization_2012,Krutyanskiy_interface_2017,Ikuta_QFC_atoms_2018,Bock_QFC_ions_2018,Maring_QFC_2018,Krutyanskiy_interface_2019}.\\
Concerning the very popular polarisation observable, a high-efficiency and versatile up-converter from telecom to the quantum memory band is still missing. One attempt based on three-wave mixing in nonlinear crystals showed up-conversion from 810\,nm to 532\,nm, but the reported conversion efficiency is $\sim 0.04\%$~\cite{ramelow_polarization_2012}. Down-conversion experiments showed however, that higher efficiencies in the few 10 percent range should be feasible~\cite{Krutyanskiy_interface_2017,Ikuta_QFC_atoms_2018,Bock_QFC_ions_2018,Maring_QFC_2018,Krutyanskiy_interface_2019}.

Our work provides a decisive step forward in this regard. We demonstrate a novel quantum interface in which single photons are converted via sum frequency generation (SFG) in nonlinear crystals from 1560\,nm to 795\,nm.
Key features of our approach are a high conversion efficiency and the preservation of polarisation quantum states with high fidelity.
We choose a design based on nonlinear guided-wave optics to increase robustness and facilitate integration into existing standard systems.
Furthermore, our device is essentially noise-free over a spectral bandwidth compatible with stationary quantum systems based on hot and cold atomic ensembles, as well as solid state quantum memories~\cite{Lvovsky_memory_2009}.
Therefore, our work represent a significant step towards interconnecting the pillars of quantum communication, storage, and computation.
\section{Setup}

\begin{figure}[t!] 
\begin{center}
\includegraphics[width=1\columnwidth]{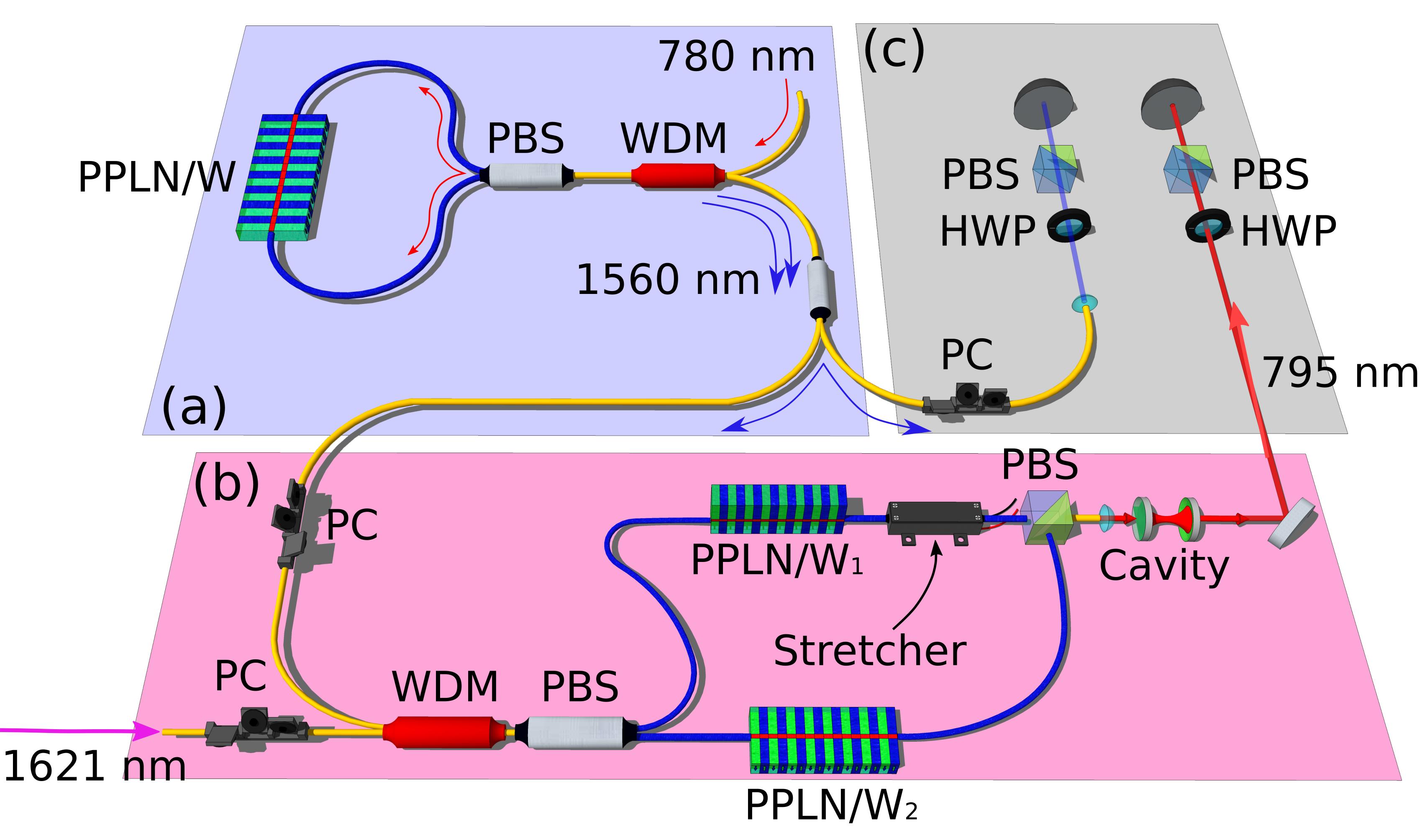} 
\caption{Experimental scheme. (a) Polarization entangled photon pair source based in a fiber-based nonlinear Sagnac loop. Photons pairs are created by SPDC into a type-0 PPLN/w at the degeneracy wavelengths 1560\,nm. (b) Polarization-coherent up-conversion setup. Signal photons (1560\,nm) and pump laser (1621\,nm) are combined into a nonlinear fiber-based MZI. The horizontally (vertically) polarization components of the input are up-converted in the top (bottom) waveguide to a horizontally (vertically) photon, i.e. mode, at 795\,nm. (c) Polarization analysis and Bell state measurements apparatus. The entangled pair, composed of the 1560\,nm and 795\,nm photons, are sent to two polarization state analyzers. The polarization rotation is implemented using two HWP optimized at their corresponding wavelengths and projected onto a beam splitter. The photons are fiber-coupled and detected by single photon counting modules permitting the registration of coincidence events.}
\label{Setup}
\end{center}
\end{figure}

The physical realisation of our quantum interface is depicted in \figurename~\ref{Setup}.
The target is to convert single photons from 1560\,nm to 795\,nm via SFG with a 1621\,nm pump laser.
Single photons and pump laser are combined into the same fibre using a standard telecom wavelength division multiplexer (WDM).
Conversion takes place in two 3.8\,cm long periodically poled lithium niobate waveguides (PPLN/W$_{1,2}$).
One PPLN/W is placed in each arm of a Mach-Zehnder type interferometer made of polarisation maintaining fibres and polarising beam splitters (PBS$_{1,2}$) at the in- and output.
PBS$_1$ splits up light into horizontally and vertically polarised components, subsequently propagating in the upper and lower arms, respectively. After wavelength conversion in the PPLN/Ws, the polarisation components are recombined into the same spatial mode at PBS$_2$.
In both crystals, we choose to exploit the type-0 interaction which is associated with the largest obtainable nonlinear coefficient.
Note that this interaction necessitates that all light fields are vertically polarised inside the crystal~\cite{Tanzilli_ppln_2002}, which is why we rotate PPLN/W$_1$ by $90^{\circ}$ around the light propagation axis.

To ensure that the wavelength conversion process does not deteriorate the polarisation state between the input and output photon, two requirements need to be satisfied~\cite{Krutyanskiy_interface_2017,Ikuta_QFC_atoms_2018,Bock_QFC_ions_2018,Maring_QFC_2018,Krutyanskiy_interface_2019}. First, the conversion efficiency in both arms needs to be equalised; second, the optical phase difference between upper and lower arms needs to be zero to avoid polarisation state rotation.
We fulfil the first condition by rotating the polarisation of the 1621\,nm laser power with a polarisation controller (PC) until each PPLN/W receives $\sim 100\rm\,mW$ of optical power.
The second condition necessitates an active interferometer phase stabilisation system. We implement it by \textit{recycling} the spurious emission at 810\,nm originating from residual frequency doubling of the 1621\,nm pump laser. We use a dichroic mirror to separate this light from the desired photons at 795\,nm. After projecting the 810\,nm photons into the diagonal basis, we observe phase-dependent interference fringes. A piezoelectric fibre stretcher in one arm of the interferometer is then used to maintain the phase stable.

Conversion devices based on PPLN show generally strong and broadband Raman scattering at $\sim 250\rm\,cm^{-1}$ and $\sim 630\rm\,cm^{-1}$~\cite{Pelc_up_2011,Kaiser_tele_2015}.
The resulting photonic noise is detrimental for quantum applications based on single photons and usually necessitates several filtering stages.
In our experiment, we actually choose deliberately to operate in a worst-case scenario to demonstrate the suitability of our filtering stage.
In PPLN, the first anti-Stokes emission peak of the 1621\,nm pump laser is situated at $\sim 1560\rm\,nm$, \textit{i.e.} overlaying with the wavelength of the photons that we want to convert.
To filter out Raman noise, we use a home-made hemispherical cavity at 795\,nm with a free spectral range of 150\,GHz. This way, we ensure that there is only one transmission peak within the spectral conversion bandwidth of our PPLN/Ws ($40 - 44\rm\,GHz$). We choose a cavity transmission bandwidth of $\sim 250\rm\,MHz$ to be compatible with the absorption of quantum systems based on atomic vapours and solid state quantum memories~\cite{Lvovsky_memory_2009}.
Thanks to the narrow transmission bandwidth, wavelength converted anti-Stokes photons within the spectral conversion bandwidth are strongly suppressed by about two orders of magnitude. We further use a 795\,nm reference laser to implement an active cavity stabilisation system with a temporal duty cycle of 6\%. Finally, before coupling light back into an optical fibre, we employ a $800\pm 20\rm\,nm$ bandpass filter to suppress light at 540\,nm originating from parasitic frequency tripling inside the PPLN/Ws.

\section{Results}

In a first step, we perform a semi-classical characterisation of our interface device. For this, we simulate a source of diagonally polarised single photons by carving 10\,ns long pulses out of a 1560\,nm continuous wave laser.
\begin{figure}[t] 
\centering
\includegraphics[width=1\columnwidth]{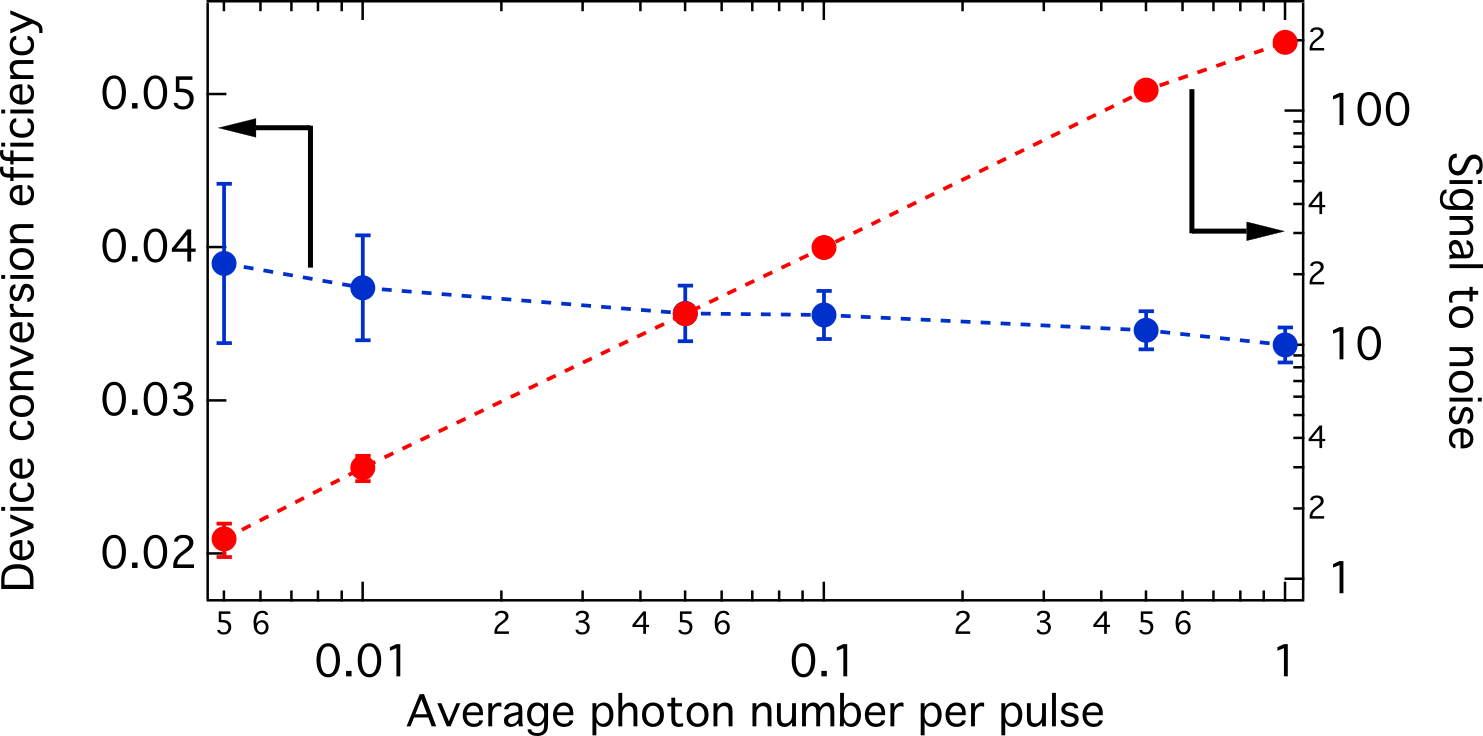} 
\caption{Device conversion efficiency $\eta$ (blue dots) and associated SNR (red dots) as a function of the average number of photons per laser pulse $\bar{n}$.
Dashed lines are guides to the eye.}
\label{eff_SNR}
\end{figure}
Via a digital attenuator, we then adjust the average number of photons per pulse in the range of $\bar{n} = 0 - 1$.
After the interface, photons arriving within a 10\,ns window are detected using a silicon single photon detector (SPD) with 50\% efficiency and a free-running dark count rate of $\sim 120\rm\,s^{-1}$.\\
\figurename~\ref{eff_SNR} shows the wavelength conversion efficiency of the full device ($\eta$) and the obtained signal-to-noise ratio (SNR) as a function of $\bar{n}$.
From the data, we infer $\eta = 3.6 \pm 0.2 \%$, which is about two orders of magnitude higher than demonstrated previously~\cite{ramelow_polarization_2012}.
To infer how our efficiency could be further improved, we analyse the optical loss of our setup. In total, we measure 8.8\,dB, broken down as follows: 3\,dB from the WDM to the PPLN/Ws; 3.7\,dB from PPLN/Ws to the cavity; 1.1\,dB inside the cavity including its stabilisation system; and 1\,dB from cavity into the optical fibre towards the SPD. Thus, we estimate the internal conversion efficiency of the PPLN/Ws to be $27.3 \pm 1.5\%$.
Concerning SNR measurements, our filtering stage proves to be effective as we achieve high levels even at low $\bar{n}$. From a linear fit to the recorded data, we obtain SNR$=243(1) \cdot \bar{n}$.

In a second step, we perform single photon polarisation state tomography to demonstrate that our quantum interface does not alter the quantum state~\cite{James_2001_measurement_of_qubits,Altepeter_Photonic_state_Tomo_2005,Bayraktar_2016_polarization_tomography}.
Setting $\bar{n} = 0.5$ we generate the six basis polarisation states $|H \rangle, |V \rangle, \tfrac{ |H \rangle + |V \rangle }{\sqrt{2}} \equiv |D \rangle, \tfrac{  |H \rangle - |V \rangle }{\sqrt{2}} \equiv |A \rangle, \tfrac{ |H \rangle + i |V \rangle }{\sqrt{2}} \equiv |R \rangle, \tfrac{|H \rangle - i |V \rangle}{\sqrt{2}} \equiv |L \rangle$ at 1560\,nm with fidelities of $\sim 0.99$. Here, $|H \rangle$ and $|V \rangle$ denote horizontally and vertically photon polarisations, respectively.
\begin{figure}[t] 
\centering
\includegraphics[width=1\columnwidth]{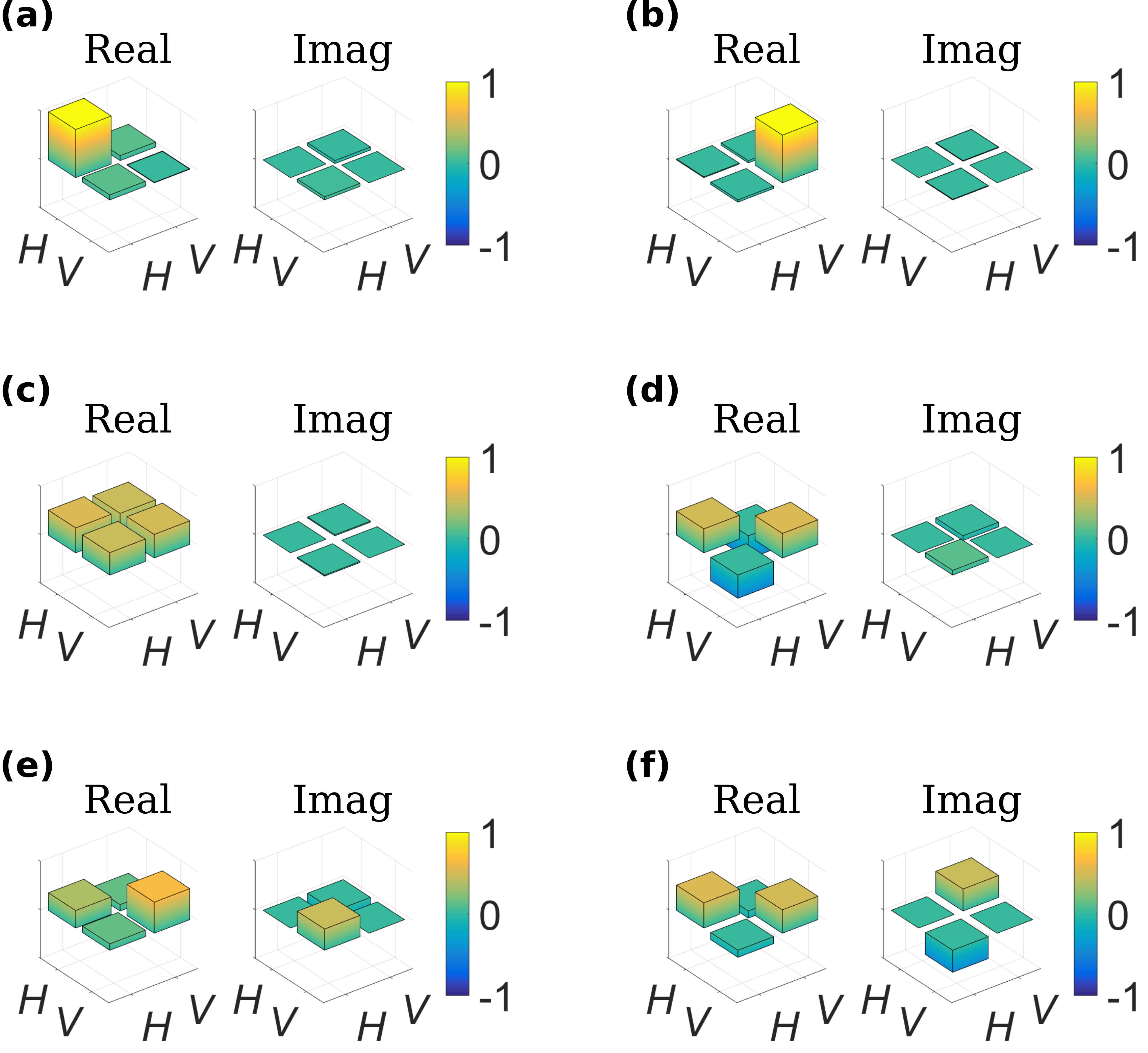} 
\caption{Real and imaginary parts of the single photon polarisation state tomography after wavelength conversion to 795\,nm.
The 1560\,nm input states are (a) $|H \rangle$, (b) $|V \rangle$, (c) $|D \rangle$, (d) $|A \rangle$, (e) $|R \rangle$, and (f) $|L \rangle$, respectively.}
\label{single_photon_tomo}
\end{figure}
These photons are then wavelength converted and their polarisation state is measured using a quarter wave plate (QWP), a half wave plate (HWP), and a PBS.
\figurename~\ref{single_photon_tomo} shows the detected quantum states at 795\,nm.
From the data, we compute quantum state fidelities of $\mathcal{F}_{|H \rangle}=0.98$, $\mathcal{F}_{|V \rangle}=0.98$, $\mathcal{F}_{|D \rangle}=0.96$, $\mathcal{F}_{|A \rangle}=0.98$, $\mathcal{F}_{|R \rangle}=0.94$, and $\mathcal{F}_{|L \rangle}=0.95$, with typical error bars of $\pm 0.01$. The origin of the fidelity decrease will be discussed later in this section.

In the final step, we proceed to wavelength conversion of polarisation entangled photons to demonstrate the polarisation-insensitivity of our device in a universal manner~\cite{Tanzilli_Interface}.
Here, we use a previously developed source based on a nonlinear Sagnac interferometer (see \figurename~\ref{Setup}\textbf{a})~\cite{Vergyris_sagnac_2017}. Photon pairs are generated at 1560\,nm in the maximally entangled polarisation Bell state $|\psi \rangle = \left( |H \rangle |H \rangle + |V \rangle |V \rangle \right) / \sqrt{2}$ with an initial fidelity of $\mathcal{F}_{\rm i} = 0.989 \pm 0.002$.
After the source, photon pairs are split up probabilistically at a fibre beam-splitter.
The photon that is not wavelength converted is filtered down to a spectral bandwidth of 500\,MHz with a phase-shifted fibre Bragg grating to roughly match the bandwidth of the noise-reduction filter cavity (250\,MHz). The photon's polarisation state is subsequently detected using a half-wave plate (HWP), a PBS, followed by a superconducting nanowire single photon detector (SNSPD, IDQ281) with 50\% efficiency and a free-running dark count rate of $250\rm\,s^{-1}$. The other photon is sent to the interface, wavelength converted, and its polarisation state is measured using a HWP, PBS, and a SPD. Both detectors are connected to a time-to-digital converter, allowing to infer two-photon coincidence events.
\begin{figure}[t] 
\centering
\includegraphics[width=1\columnwidth]{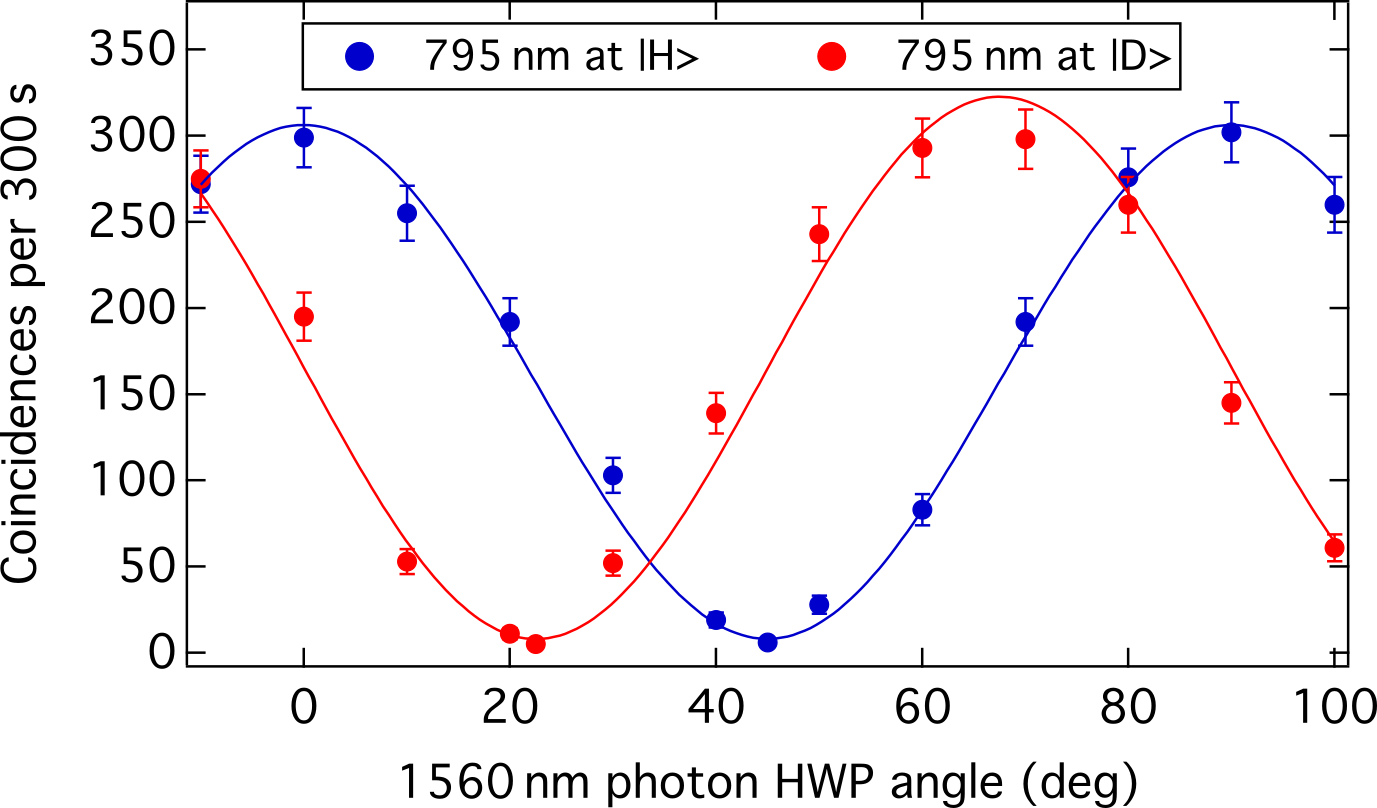} 
\caption{Two photon coincidences as a function of the 1560\,nm HWP angle.
The polarisation state of the 795\,nm is being projected onto $|H \rangle$ (blue dots) and $|D \rangle$ (red dots), respectively.
Lines are sinusoidal fits to the data from which we extract fringe visibilities of $\mathcal{V}_{|H \rangle} = 94.9 \pm 0.2\%$ and $\mathcal{V}_{|D \rangle}=95.2 \pm 0.2\%$, respectively.}
\label{Bell_violation}
\end{figure}
\figurename~\ref{Bell_violation} shows the two photon interference fringes acquired when the polarisation state of the 795\,nm is projected onto $|H \rangle$ and $|D \rangle$, respectively, and the polarisation state of the non-converted photon is rotated continuously.
We measure net interference fringe visibilities of $\mathcal{V}_{|H \rangle} = 94.9 \pm 0.2\%$ and $\mathcal{V}_{|D \rangle}=95.2 \pm 0.2\%$. From the average visibility $\bar{\mathcal{V}}$, we estimate a quantum state fidelity of $\mathcal{F}_{\rm net} = \left( \bar{\mathcal{V}} + 1\right) / 2 = 0.976 \pm 0.001$.
Without subtracting noise terms, we infer a raw fidelity of $\mathcal{F}_{\rm f, raw} = 0.945 \pm 0.001$.
The total 5.4\% drop in fidelity compared to $\mathcal{F}_{\rm i}$ has three main origins. From $\mathcal{F}_{\rm f, net}$, we conclude that 1.3\% degradation is due to non-optimal setup alignment, \textit{i.e.} non-equilibrated conversion efficiencies between the two PPLN/Ws and phase fluctuations inside the interferometer.
By performing a measurement with the 1621\,nm pump laser being switched off, we infer that 1.6\% loss in fidelity is due to detector dark counts (effectively 0.02 noise photons per second after coincidence gating).
The remaining 1.5\% come from photonic noise due to Raman scattering.

Our current quantum state transfer fidelity is therefore $\mathcal{F}_{\rm trans} = \mathcal{F}_{\rm f,raw} / \mathcal{F}_{\rm i} = 0.956 \pm 0.002$. Through technical improvements, such as better setup alignment and employing ultra low noise detectors, an optimal fidelity $\mathcal{F}_{\rm trans}^* = 0.985$ could be reached.
Finally, we mention again that we deliberately chose to operate our experiment in worst-case scenario regarding photonic noise. For \textit{optimal} wavelength combinations, \textit{i.e.} conversion from 1530\,nm to 795\,nm, photonic noise is almost two orders of magnitude lower~\cite{Pelc_up_2011}, such that near unit transfer fidelities are definitely realisable. Alternatively, the filter cavity could be removed, thus increasing the the full device conversion efficiency by 1.1\,dB.

\section{Conclusion and discussion}

We have demonstrated polarisation state preserving quantum frequency conversion from 1560\,nm to 795\,nm.

In our setup, we achieved a total device conversion efficiency of 3.6\%, which could be increased by a factor $3$ to $4$ with a more powerful 1621\,nm pump laser. Another twofold improvement is feasible through reducing optical losses, \textit{e.g.} by splicing all fibres and employing custom optical filters and dichroic mirrors.

Although we chose to operate in a worst-case scenario concerning photonic noise induced by Raman scattering, we show excellent signal-to-noise thanks to a filter cavity stage.
This allowed us to wavelength convert one photon out of an entangled pair with more than 95\% fidelity, and pathways towards achieving near-unit fidelities have been outlined.

Since our device is almost exclusively based on fibre and guided-wave technology, integration into existing systems is greatly simplified. In this perspective, we mention that essentially no efforts have to be made to establish and maintain a good mode overlap between the single photons and the pump laser which further increases robustness. Further stability improvements could be made with a fully integrated QFC device, and promising efforts in this direction have recently been demonstrated, \textit{e.g.} on-chip combination of PPLN/Ws and PBS~\cite{Sansoni_PBS_2017} or high-rejection spectral filtering stages~\cite{Perez_BraggFilter_2017}.

We also stress that our particular choice of wavelengths and the adapted transmission bandwidth of the filter cavity makes our setup ready to be used in a quantum repeater based network. Here, the idea is to distribute 1560\,nm entangled photons in a fibre network, convert them to 795\,nm, and store them in a (hot) rubidium atom based quantum memory. Cold atom memories with absorption bandwidth in the few MHz range can also be addressed, however, the filter cavity bandwidth should be reduced in this case to guarantee a high signal-to-noise ratio.

By combining our work with recently demonstrated downconversion interfaces~\cite{Krutyanskiy_interface_2017,Ikuta_QFC_atoms_2018,Bock_QFC_ions_2018,Maring_QFC_2018}, we are now a significant step closer to a universal quantum network based on frequency conversion back and forth between quantum systems based on light and matter.

\section*{Funding}

Foundation Simone \& Cino Del Duca; European Commission (FP7-ITN PICQUE project, grant agreement No 608062); Agence Nationale de la Recherche (ANR) (e-QUANET grant ANR-09-BLAN-0333-01, CONNEQT grant ANR-EMMA-002-01, INQCA grant ANR-14-CE26-0038, and SPOCQ grant ANR-14-CE32-0019, Quantum$@$UCA grant ANR-15-IDEX-01); iXCore Research Foundation; French government (Universit\'e C\^ote d'Azur UCA-JEDI project); European Union and R\'egion PACA (OPTIMAL project via Fond Europ\'een de D\'eveloppement R\'egional, FEDER).

\section*{Disclosures}
The authors declare that there are no conflicts of interest related to this article.

\end{document}